\documentclass{iopart}

\usepackage{amsmath} 
\usepackage{amssymb} 
\usepackage{graphicx}
\usepackage{xspace}
\newcommand{\p}{\partial} 

\newcommand{\fra}{\displaystyle \frac} 
\newcommand{\inte}{\displaystyle \int}

\newcommand{\tp}{\tilde{\psi}}
\newcommand{\bp}{\bar{\rho}}

\newcommand{\tphi}{\tilde{\phi}}

\newcommand{\tj}{\tilde{j}}

\newcommand{\gk}{\Gamma_k}
\newcommand{\dtt}{\tilde{\partial}_s}
\newcommand{\dimz}{\bar{Z}_k}
\newcommand{\dimx}{\bar{X}_k}

\newcommand{\anz}{{\it Ansatz}\xspace}
\newcommand{\nequ}{non-equilibrium\xspace}
\newcommand{\npt}{non-perturbative\xspace}

\newcommand{\pk}{\partial_k}
\newcommand{\expo}[1]{e^{\displaystyle #1}}
\newcommand{\demi}{\frac{1}{2}}

\newcommand{\clw}{{\cal W}_k}

\newcommand{\tJ}{\tilde{j}}
\newcommand{\rk}{\hat{R}_k}
\newcommand{\xg}{\bold x}
\newcommand{\yg}{\bold y}

\newcommand{\dJ}[1]{\hat{\delta}_{{\cal J}_{#1}}}
\newcommand{\derw}{\hat{\cal W}_k^{(2)}}
\newcommand{\dP}[1]{\hat{\delta}_{{\Psi}_{#1}}}
\newcommand{\derg}{\hat{\Gamma}_k^{(2)}}
\newcommand{\sca}{.}

\begin{document}

\title{Non-perturbative Approach to  Critical Dynamics}
\author{L\'eonie Canet}
\address{Service de Physique de l'\'Etat Condens\'e, CEA  Saclay, 91191
~Gif-sur-Yvette Cedex,~France \\  Laboratoire de Physique et Mod\'elisation des Milieux Condens\'es,  CNRS, 25 avenue des Martyrs,
BP 166 - 38042  Grenoble Cedex, France}

\author{Hugues Chat\'e}
\address{Service de Physique de l'\'Etat Condens\'e, CEA  Saclay, 91191
~Gif-sur-Yvette Cedex,~France}

\begin{abstract}
 This paper is devoted to a non-perturbative renormalization group 
(NPRG) analysis of Model A, which stands as a paradigm for the study of critical dynamics. 
 The NPRG formalism  
  has appeared as a valuable theoretical tool to investigate 
 \nequ critical phenomena, yet the simplest --- and nontrivial --- 
 models for critical dynamics have never been
  studied using  NPRG techniques. 
 In this paper we focus on Model A taking this opportunity to provide a pedagological introduction
 to  NPRG methods for dynamical problems in statistical physics.
   The dynamical exponent $z$  is computed 
 in $d=3$ and $d=2$  and is found in close agreement with results from other methods.

\end{abstract}

\date{\today}

\section{Introduction}
\label{secIntro}

The understanding of \nequ critical phenomena stands as one of the major
 challenges of statistical physics. Systems far from thermal equilibrium 
 are omnipresent in nature:
   slow relaxation or  external driving forces
  tend to prevent real systems  from ever reaching their equilibrium distribution.
  Behavior out-of-equilibrium is far richer than at equilibrium, and many
 intriguing scaling phenomena, such as self-organized criticality (emergence of scaling 
 without fine-tuning of a  control parameter) \cite{bak87},
  or phase transitions
 between \nequ stationary states \cite{hinrichsen00,odor04},
  have been observed for long. 
 However, despite the considerable achievements of equilibrium statistical 
 physics, 
 the theoretical comprehension of \nequ critical phenomena remains much poorer.
  The renormalization group (RG), which has appeared as a cornerstone
   to explain universality in equilibrium continuous phase transitions,
  has  also allowed some  breakthroughs
 out-of-equilibrium \cite{tauber05}. Nonetheless, many \nequ phenomena 
 remain out of range of perturbative approaches because of
 large coupling constants or because the interesting dimensions
 lie  far from the critical one.
 Further theoretical progress 
  out-of-equilibrium  is hindered by the lack of analytical
  tools  to handle the corresponding models. 

 Recently, a novel  approach -- namely the \npt renormalization group (NPRG) --
  has been  proposed to investigate (\nequ)
  reaction-diffusion processes \cite{canet04a}. It  has allowed 
 to overcome the perturbative limitations  and to 
  gain physical insights into 
  models such as   branching and annihilating random
 walks --- which are reviewed in \cite{canet05b}. For instance, these studies have captured 
    \npt effects that essentially determine
 the phase diagram of some systems \cite{cardy96,cardy98,canet04a,canet04b}
  and have unveiled
  a genuinely \npt fixed point governing the phase transition
 belonging to the so-called parity-conserving universality class \cite{canet05}. 
 A valuable advantage of this method
 is that it gives a unified description
 of a model:   the very same 
 equations  enable one  to probe 
 any dimensions or coupling regimes, including non-universal features. 
 Hence, the NPRG appears as a powerful tool to investigate
 \nequ systems. 

However most readers are still largely unfamiliar with these techniques,
 though they have been introduced more than a decade ago for systems 
 in  equilibrium \cite{wetterich93,morris94}.
 The aim of this contribution is to give a
 brief but pedagological introduction
 to the NPRG methods for \nequ systems. 
 For this purpose, Model A stands as one of the simplest
 -- yet far from being trivial -- dynamical models and it has 
  never been studied within the NPRG framework so far,
  which we  remedy with the present work.
 We adopt  a practical viewpoint
 and put a particular emphasis  on
  the discussion of the 
 device of a (\npt) approximation scheme.

The paper is organized as follows. In Section  \ref{secModA} we briefly recall
 the definition of Model A and its critical properties. In Section \ref{secNPRG},
 the principles and the construction of the NPRG are outlined
 for generic \nequ systems and
 specialized in Section \ref{secAnz} to Model A. The corresponding
 \npt flow equations are derived in Section \ref{secFlow} and their numerical
 integration is dealt with in Section \ref{secNum}.
 The results are eventually discussed in Section \ref{secRes} and followed by 
 a brief summary in Section \ref{secConcl}.

\section{Model A and critical dynamics}
\label{secModA}
 
The purely dissipative relaxation  of a non-conserved  field $\phi$
   can be described by the Langevin equation:
\begin{equation}
\p_t \phi(x,t) = - D \, \displaystyle \frac{\delta {\cal H}[\phi]}{\delta \phi(x,t)} + \eta(x,t)
\label{lang}
\end{equation}
where $D$ denotes a constant and uniform relaxation  rate and ${\cal H}$  the usual
  Landau-Ginzburg-Wilson Hamiltonian. 
 On approaching a critical point,  the relaxation time of the order parameter
  starts diverging, which reflects  the critical slowing down of the dynamics.
  The Langevin equation is a `mesoscopic' description of the system
 which exploits  the associated decoupling of  time scales:
  the 
 order parameter, represented by the  (coarse-grained) field $\phi$,
  relaxes much slower than  all  the other microscopic 
  degrees  of freedom, 
which can hence be modeled by a  stochastic
 Gaussian noise   $\eta$ with zero mean and
  variance
\begin{equation}
\big\langle \eta(x,t)\eta(x',t')\big\rangle = 2 D\, k_B T\, 
\delta^d(x-x')\,\delta(t-t').
\label{var}
\end{equation}
 The strength of the noise  is fixed by  the Einstein relation which ensures
  that the system acquires its equilibrium distribution at long time.
  We here focus on the
 case of  a scalar order parameter with Ising symmetry, 
 described by the Hamiltonian
\begin{equation}
 {\cal H}[\phi] =  \inte d^d x \left(\fra{1}{2} [\nabla \phi(x)]^2 + U(\phi) -h(x) \phi(x) \right),
 \hbox{\hspace{1cm}} U(\phi)= \fra{r}{2} \phi^2 + \fra{u}{4!}  \phi^4 .
\label{ham}
\end{equation}
The Langevin equation (\ref{lang}) corresponds  to a  Glauber dynamics
 for the Ising spin and defines Model A in
  the classification by Halperin and Hohenberg \cite{hohenberg77}. 
Besides the equilibrium critical exponents $\nu$ and $\eta$,
  the critical dynamics  is characterized  by the dynamical exponent $z$
  that describes the divergence of the
  relaxation time $\tau\sim \xi^z\sim |T-T_c|^{-z\nu}$
 near the temperature $T_c$ of the second-order phase transition.

\subsection*{Time Reversal Symmetry}

In the long-time limit after the initial perturbation,  the system is
expected to become  time translational invariant (TTI) and  the time-reversal symmetry (TRS) to hold.
   TRS then yields the fluctuation-dissipation theorem (FDT) which
 linearly relates the two-point correlation function 
 $C(x-x',t-t')= \langle \phi(x,t) \phi(x',t')\rangle$
  with the response function 
$R(x-x',t-t') = \delta\langle \phi(x,t)\rangle/\delta h(x',t')$ following:
\begin{equation}
 R(x-x',t-t')= -\theta(t-t')\fra{1}{T} \p_t C(x-x',t-t').
\label{fdt}
\end{equation}

In the early stages of the relaxation process,  the system is generally 
 not TTI
  such that  $R$ and $C$ may depend on both $t$ and $t'$
 and FDT may not hold. The 
 fluctuation-dissipation ratio  $T R/\p_t C$  
 becomes of particular interest to characterize the violation of FDT
 and the associated ageing phenomena (see \cite{calabrese05} for a recent review
 and references therein).
 For our \npt study,  we rather focus
  on the stationary dynamics where the system 
 satisfies TRS.

\subsection*{Field Theory}

Any RG treatment starts out from  a field theory.
Upon introducing a Martin-Siggia-Rose response field $\tphi(x,t)$ \cite{martin73},
  one can average over the Langevin noise $\eta$ and cast the 
 stochastic equation (\ref{lang}) into a dynamic functional \cite{janssen76,dedominicis76}:
\begin{equation}
{\cal S}[\phi,\tphi] = \inte d^d x\,dt \, \left\{ \tphi\left[\p_t
  \phi +   D \, \fra{\delta {\cal H}[\phi]}{\delta \phi}\right] \, - D \tphi^2  \right\}
\label{bareact}
\end{equation}
(where $k_B T$ has been set to unity).
Correlation and response functions can then be expressed as functional averages with the weight $\exp (- {\cal S}[\phi,\tphi])$. 

In this equivalent field theoretical formulation,  TRS 
 can be conveniently  expressed as an invariance of the action (\ref{bareact})
 under a specific field transformation, as stressed in \cite{biroli05}. 
 This transformation writes 
\begin{equation}
\left\{
\begin{array}{l l l}
\phi & \to & \phi \\
\tphi & \to & \tphi - \fra{1}{D}\p_t\phi.
\end{array}
\right.
\label{trs}
\end{equation}
Indeed, one can straigthforwardly check that 
after performing a time inversion $t\to -t$ in
 (\ref{bareact}) which switches the sign of the kinetic term 
 $\tphi \,\p_t \phi$, the field transformation (\ref{trs}) 
 yields  additional contributions  
  from the latter term and from the noise term $D \tphi^2$ that cancel out.
  Besides, the transformation of the Hamiltonian part under  (\ref{trs}) produces an additional term 
 $\propto \p_t \phi \,{\delta {\cal H}}/{\delta \phi}$, which  vanishes upon time integration 
  in the stationary regime. 
 We shall rely in the following on this simple expression of  
 TRS  to ensure that this invariance is preserved
 within the \npt formulation.

\section{The NPRG formalism in \nequ statistical physics}
\label{secNPRG}
 
The NPRG formalism relies on Wilson's RG idea \cite{wilson74}, which consists in building  a sequence of scale-dependent effective Hamiltonians,
that interpolate smoothly between the short-distance physics at the
(microscopic) scale $k=\Lambda$ and the long-distance physics at the scale $k=0$,
 through progressively averaging over fluctuations.
Rather than expressing --- as in the original Wilsonian formulation --- the flow of effective Hamiltonians for the slow modes,
one can work out the flow of effective `Gibbs free energies' $\gk$ for the rapid ones, following \cite{tetradis94,berges02}. $\gk$ thus only includes fluctuation modes with momenta $|q| \geq k$. At the scale $k=\Lambda$, no  fluctuation is yet  taken into account
and $\Gamma_{\Lambda}$ coincides with the microscopic action ${\cal S}$ \cite{tetradis94}, while at  $k=0$, all  fluctuations are integrated out and $\Gamma_{0}$ is the analogue of the Gibbs free energy $\Gamma$ at thermal equilibrium, in that it encompasses the long-distance and long-time
 properties of the system.
Hence, to construct $\gk$, one needs at a given scale $k$ to suppress the slow modes with momentum $|q|<k$. 
This is achieved by adding to the original action (\ref{bareact}) a scale-dependent term  \cite{berges02,canet04a,canet04c} which is quadratic in the fields (so as to affect the propagator of the corresponding modes):
\begin{equation}
\Delta {\cal S}_k[\phi,\tphi] = \fra{1}{2} \int_{x,t} [\phi(x,t),\tphi(x,t)] \,\hat{R}_k(\nabla^2,\p_t) \,^t[\phi(x,t),\tphi(x,t)],
\label{mass}
\end{equation}
where $\hat{R}_k$ is a  symmetric $2 \times 2$  matrix of elements $R_k^{i j}$ ($i,j=1,2$). 
 These elements (so-called cutoff functions) will be specified in the following, but 
 their general properties can be already stressed.
  In order to achieve the renormalization procedure outlined 
 above,  these cutoff functions must behave at fixed $k$ as
  $R_k^{i j }\sim k^2$ (in Fourier space) 
 for small momenta $|q| \leq k$
 --- so that the slow fluctuation modes acquire a `mass' $k^2$ and decouple.
  On the other hand, $R_k^{i j}$ must vanish for large momenta $|q| \geq k$
  --- so that the rapid modes remain unaltered and contribute to the functional averages
 with weight $\exp(-{\cal S}_k)$.
 Besides, the  additional constraints
\begin{equation} 
 \lim_{k\to 0} R_{k}^{i j} = 0,\hbox{\hspace{1cm}}
  \lim_{k\to \Lambda} R_{k}^{i j}=\infty \hbox{\hspace{1cm} at fixed $q$}
\end{equation}
 must be satisfied in order to enforce 
 the correct asymptotic behaviors at the  scales 
  $k=\Lambda$ and $k= 0$, respectively $\Gamma_{k=\Lambda}\sim
{\cal S}$ and $\Gamma_{k=0}=\Gamma$ \cite{berges02,canet04a,canet04c}.

With the additional term (\ref{mass}) the `partition
functions' 
\begin{equation}
{\cal Z}_k[j,\tj]= \int {\cal D}\phi\, {\cal D}i \tphi\, \exp(- {\cal S}- \Delta {\cal S}_k +  \int j \phi +  \int \tj\tphi) 
\label{calz}
\end{equation}
 become $k$-dependent.
 Finally, the effective  $\gk$ which is the central object
 of the NPRG procedure 
  is defined as  the (modified) Legendre transform of   
$\log {\cal Z}_k[j,\tj]$:
\begin{equation}
\Gamma_k[\psi,\tp] +\log {\cal Z}_k[j,\tj]= 
\int j \psi +\int \tj \tp -\Delta {\cal S}_k[\psi,\tp].
\label{legendre}
\end{equation}
 $\gk$ is a functional of the conjugate fields
$\psi=\delta \log {\cal Z}_k/\delta j$ and $\tp=\delta \log {\cal Z}_k/\delta
\tj$.
 The additional term $\Delta {\cal S}_k$ in Eq. (\ref{legendre}) is necessary to set
  the proper microscopic behavior  at $k=\Lambda$: $\Gamma_{k=\Lambda}\sim
{\cal S}$ \cite{berges02}. 
The RG flow of $\Gamma_k$ under an infinitesimal change of
the scale $k$ --- or rather $s=\log(k/\Lambda)$ ---  is governed by an exact
  functional differential equation  \cite{berges02,canet04c} (which is  derived in the Appendix):
\begin{equation}
\partial_s \Gamma_k = \frac{1}{2} {\rm Tr} \int_{q,\omega} \partial_s \hat{R}_k \left(\hat\Gamma_k^{(2)} + \hat{R}_k\right)^{-1}.
\label{dkgam}
\end{equation}
In this equation, $\hat\Gamma_k^{(2)}[\psi,\tp]$ is the $2\times
2$ matrix of second derivatives of $\Gamma_k$ with respect to (wrt) $\psi$ and
$\tp$ and $[\hat\Gamma_k^{(2)} + \hat{R}_k]^{-1}$  hence embodies 
 the full (functional) propagator
 associated with the effective theory ${\cal S}_k+ \Delta{\cal S}_k$.

 Obviously, Eq.~(\ref{dkgam}) cannot be solved exactly
and one has to resort to some approximations \cite{berges02}. 
  However, as the approximations used do not rely on the smallness of a parameter
 (see next section),
  the approach remains \npt in essence. In particular,  
 it is not confined to weak-coupling regimes or 
 to the vicinity of critical dimensions and
   is therefore suitable to overcome the limitations of 
 perturbative RG schemes.

\section{NPRG for Model A}
\label{secAnz}

To exploit the exact flow equation (\ref{dkgam}), one has to 
 device an approximation scheme. This scheme is  based on the construction of
  an \anz for $\Gamma_k$ which does not spoil the \npt features
 of the exact equation --- and which can be systematically 
 enlarged. The formulation of this \anz
  relies on the physics one wishes 
 to probe, that is some basic physical insights are necessary. 
 The most  common truncation consists  in
 expanding  $\Gamma_k$ in powers of  gradients \cite{tetradis94} and time derivatives.
 The accuracy and convergence of this approximation scheme have been
 thoroughly studied in the equilibrium context 
 and have shown that quantitatively
  reliable results can already be obtained at the leading order ($\nabla^2$) \cite{berges02}.
 For instance, for the three-dimensional Ising model, NPRG calculations yield
  for the critical exponents $\nu=0.628$ and $\eta=0.0443$ at order $\nabla^2$ 
 \cite{berges02,canet03b}, 
  and $\nu=0.632$ and $\eta=0.033$ at order $\nabla^4$ \cite{canet03b}
 which are in close agreement with the 6-loop results $\nu=0.6304(13)$ and $\eta=0.0335(25)$ 
  \cite{guida98}.
  Another useful approximation scheme is the field expansion of  $\Gamma_k$.
  This truncation has the advantage of preserving 
  the momentum structure of higher order vertices but it approximates the functional
 structure of the effective potential \cite{berges02}.
  The derivative expansion is best  appropriate 
  for the study of critical physics which is conveyed by 
 the large-distance ($q\to 0$)  and long-time ($\omega \to 0$) modes.
 We hence adopt here this approximation scheme and expand $\gk$
 at leading order in derivatives --- {\it i.e.}  only terms of order $\nabla^2$
 and $\p_t$  are retained.

\subsection*{Construction of an \anz for $\gk$ at leading order}

 The form of the \anz for $\Gamma_k$ is  dictated by the symmetries.
 Since   we consider the long-time regime where TRS holds,
   we want the \anz to  be invariant under the field transformation 
 (\ref{trs}) (where  $D$ is set to unity),
 which in turn imposes the following  structure:
\begin{equation}
\gk(\psi,\tp)=\int d^dx \, d t\, \left\{\tp\,  X^a_k(\psi)\,\p_t \psi + \tp\, X^b_k(\psi,\nabla \psi) -  X^c_k(\psi,\nabla \psi)\,\tp^2\right\}.
\label{anzinit}
\end{equation}
 No higher  powers of the response field are allowed 
 at  order $\p_t$  due to  TRS. Indeed, 
 the transformation (\ref{trs}) would connect
  a generic `noise'  term $\tp^n X_k^d$ $n>2$ to higher order kinetic terms  
$\tp^j  (\p_t \psi)^{n-j} X_k^d$, $j=0\dots n$ which are discarded at order  $\p_t$.
 Further constraints on the $X_k^i$'s, $i=a,b,c$ can be deduced from TRS in the same
   way as in Section \ref{secModA}.
 First one must have $X^c_k = X^a_k \equiv X_k$ for the additional
 contributions   generated  by the transformation (\ref{trs}) of the `noise'
  and the `kinetic' terms to cancel out.
 As for  the linear term in $\tp$, $X^b_k$ should 
 write as a (field) derivative of a functional
 $X^b_k(\psi,\nabla\psi)\equiv\delta {\cal F}_k/\delta \psi$ for its transform
   under (\ref{trs}) to 
 vanish upon time integration.  
 We naturally adopt for ${\cal F}_k$ the  usual equilibrium \anz
 at  order $\nabla^2$  for the Ising model which has been widely studied in the past
  \cite{berges02,canet03b}
\begin{equation}
{\cal F}_k[\psi] =\int d^dx \,  \left\{\fra{1}{2}\, Z_k(\psi)\, [\nabla \psi]^2 + U_k(\psi)\right\}.
\label{ising}
\end{equation}
In this \anz, 
  the functional $U_k$ embodies the effective potential
 and  the renormalization function $Z_k$ encompasses the anomalous dimension 
  of the field (see below).

 Furthermore,  at the scale $k=\Lambda$, $\Gamma_\Lambda$ must identify
  with the microscopic (bare) action (\ref{bareact}) --- up to
 the response field rescaling $\tp \to \tp/X_\Lambda$ --- 
 {\it i.e.} one has:
\begin{equation}
 X_\Lambda \equiv \fra{1}{D}, \hbox{\hspace{1cm}} Z_\Lambda \equiv 1, \hbox{\hspace{1cm}} U_\Lambda \equiv  U(\phi) \hbox{\hspace{1cm}($U(\phi)$ is defined in Eq. (\ref{ham}))}.
\end{equation}
 The  \anz for Model A at leading order finally writes
\begin{equation}
\gk(\psi,\tp)=\int_{x,t} \Big\{\tp\,
 X_k\,\p_t \psi - \tp\,\left[Z_k(\psi)\,\nabla^2\,\psi +\fra{1}{2}\,\p_\psi Z_k(\psi) [\nabla \psi]^2 \right] 
 +   \tp \,\p_\psi U_k(\psi) - X_k\, \tp^2.
\label{anz}
\end{equation}
  This \anz constitutes the basis of our work.

\subsection*{Definition of the critical exponents}

 We now discuss how  the critical exponents  
  can be computed within the NPRG approach.
 In the critical regime, $Z_k$ is  expected
 to endow a scaling form $\dimz\sim k^{-\eta_Z}$ and followingly 
 $\psi$ to behave as $\psi\sim k^{(d-2+\eta_Z)/2}$.
 The critical exponent $\eta$  hence corresponds to $\eta_Z=-\p_s \ln \dimz$ 
 at the critical point \cite{berges02}.
 Similarly,  $X_k$ is expected to scale as $\dimx\sim k^{-\eta_X}$ 
 at the critical point such that  $\omega \sim k^{2-\eta_z + \eta_X}$ according
 to a scaling analysis of  Eq. (\ref{anz}). Hence, 
  the  dynamical exponent $z$, which by definition characterizes
  the divergence of the time scale following $\omega=k^z$,
  is  given  by  $z=2-\eta_Z +\eta_X$  where $\eta_X =-\p_s \ln \dimx$.

  Notice that 
   $Z_k(\psi)$ is a  functional of  $\psi$ 
  whereas the scaling form $\dimz$  should be a mere ($k$-dependent) number.
  In general, one defines $\dimz$
   as the value of $Z_k(\psi)$  at a given point $\psi_0$, $\dimz\equiv Z_k(\psi_0)$.
  Similarly,  though $X_k$ here is not a functional, its flow equation
  depends on $\psi$, $U_k(\psi)$, $Z_k(\psi)$ and their derivatives.
  Henceforth, the notation  $\dimx$ will  mean that the corresponding 
  expressions are evaluated for $\psi=\psi_0$.
  Of course, within the {\it exact} renormalization
 flow, the critical exponents should not {\it in fine} depend on
 the choice of  $\psi_0$. However any approximation introduces a residual
  dependence and the choice of $\psi_0$ may become important.
 The advocated choice (from equilibrium studies) is
  the (running) minimum of the effective potential $U_k$ which is implicitely 
 defined by $\p_\psi U_k(\psi_0)=0$, for it possesses the best `stability' properties \cite{berges02}.

\subsection*{Cutoff matrix}

Our last discussion to complete the settings of the NPRG formalism
 for Model A concerns the choice of the cutoff matrix $\hat{R}_k$. 
 The previous symmetry requirements  obviously also apply
  for the quadratic term $\Delta {\cal S}_k$, 
 which must in particular be invariant under TRS
 (in the stationary regime).

 The minimal \npt renormalization scheme 
 consists in performing  a space 
 coarse-graining on the propagator mode $\tp \psi$, which amounts
 to considering an off-diagonal cutoff matrix $\hat{R}_k$ 
 with elements 
\begin{equation}
 R^{12}_k= R^{21}_k \equiv Q_k(q^2) \hbox{\hspace{1cm}and\hspace{1cm}} 
  R^{11}_k= R^{22}_k=0.  
\label{cutoffmat}
\end{equation}
 This  form for  $\hat{R}_k$ is the most natural extension of the equilibrium
  case --- where one introduces the  scale-dependent quadratic 
  term $\int_q \Delta {\cal H}_k = 1/2 \,\phi_{\text{-}q} \, Q_k(q^2) \, \phi_q$ 
 in the (equilibrium) partition function ${\cal Z}_k$ to achieve the splitting of the 
  fluctuation modes \cite{berges02}.
  We recall that the cutoff function $Q_k(q^2)$
  must decay fastly for large momentum modes and behave as $k^2$ 
 for slow modes as emphazised in Section \ref{secNPRG}.
  A typical cutoff function which has been widely used since it allows
  for analytical results is the $\theta$ cutoff introduced by Litim \cite{litim01}
\begin{equation}
 Q_k(q^2) \propto (k^2 -q^2) \theta(k^2 -q^2).
\label{litim}
\end{equation}

 It turns out that, even when considering the dynamics, space coarse-graining is enough 
 to achieve a proper  \npt renormalization program 
 since the frequency integrals appear to be convergent
 and need not be regularized \cite{canet04a}. The cutoff matrix  
 (\ref{cutoffmat}) has hence been adopted in all  previous \nequ
  studies and more specifically for reaction-diffusion processes \cite{canet05b}.
 
Notice that, on the other hand, 
 one could expect that a time coarse-graining on the $\psi \tp$
  propagator may improve the procedure, though
  it has never been tested. A time coarse-graining could be achieved
 by adding a frequency cutoff $R_k^{12}=\Omega_k(i \omega)$ on $\psi \tp$ modes. 
 But in this case one would have to coarse-grain the noise part correspondingly
 in order to sustain  TRS ({\it i.e.} the invariance under the
  transformation (\ref{trs}) of $\Delta {\cal S}_k$).
 This would amount to introducing an additional  cutoff  on $\tp\tp$ modes, of the form
  $R_k^{22}=-2i/\omega \Omega_k(i \omega)$.
  The properties of  this mixed regularization scheme have never been investigated as yet
 and  is  left for future work since it represents a great deal of numerical efforts.

\section{Flow equations}
\label{secFlow}

The NPRG flow equations for the renormalization functions $U_k$, $Z_k$ 
 and $X_k$ are drawn from the exact flow of $\gk$  given by  Eq. (\ref{dkgam}).
According to the \anz (\ref{anz}),   $\p_\psi U_k$ can be defined 
 by 
\begin{equation}
 \p_\psi U_k = \fra{\delta^{d+1}(0)}{(2\pi)^{d+1}}\lim_{p,\nu\to0}
 \fra{\delta \gk}{\delta \tp(p,\nu)}\Big|_{\tp=0} \label{defuk}
 \end{equation}
 where the limit of vanishing external momentum and frequency 
  $p,\nu\to0$ means that the fields are evaluated in
  uniform and stationary configurations and the prefactor just corresponds to 
 the volume of the system in Fourier space.
 Similarly, the renormalization functions $Z_k$
 and $X_k$ can be defined by:
\begin{eqnarray}
 Z_k &=& \fra{(2\pi)^{d+1}}{\delta^{d+1}(0)}\, \lim_{p,\nu\to0}\,\p_{p^2} \fra{\delta^2 \gk}{\delta \tp(p,\nu) \delta \psi(-p,-\nu)} \label{defzk}\\
   X_k &=& \frac{(2\pi)^{d+1}}{\delta^{d+1}( 0)}\,\lim_{p,\nu\to0}\,\p_{i \nu}\fra{\delta^2 \gk}{\delta \tp(p,\nu) \delta \psi(-p,-\nu)}\label{defxk}.
 \end{eqnarray}
Obviously, $X_k$ can  alternatively be defined from the noise part as $X_k \propto \delta^2 \gk/\delta \tp^2$
 for a uniform and stationary configuration. One can check that both definitions lead to the same 
 flow equation $\p_s X_k$ which in turn reflects that TRS is preserved by the NPRG flow at any scale $s$.

The flow equations of the renormalization functions $\p_\psi U_k$, $Z_k$ and $X_k$ 
are obtained by taking
  the scale derivative $\p_s$ of the expressions (\ref{defuk}),
  (\ref{defzk}) and (\ref{defxk}) respectively. It is convenient to first
 rewrite the flow equation $\p_s \gk$ (given by (\ref{dkgam})) as 
\begin{equation}
\partial_s \Gamma_k = \frac{1}{2} \dtt {\rm Tr} \ln  \left(\hat\Gamma_k^{(2)} + \hat{R}_k\right),
\label{dttgam}
\end{equation}
where $\dtt(.)\equiv \hat{R}_k {\delta (.)}/{\delta \hat{R}_k}$ only acts  
 on the $s$-dependence of the cutoff elements $R_k^{ij}$.
 It follows that the field derivatives of $\p_s \gk$ 
 admit  simple diagrammatic representations:
\begin{center}
\begin{minipage}[t]{2.7cm}
\vspace{-0.8cm}
  $ \fra{\delta \p_s\,\gk}{\delta\tp(0,0)} = \fra{1}{2} \, \dtt$
\end{minipage}
\begin{minipage}[t]{5cm}
\includegraphics[width=12mm,height=10mm]{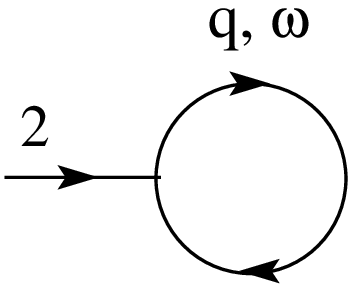}
\end{minipage}
\end{center}
\begin{center}
\begin{minipage}[t]{4.5cm}
\vspace{-1.2cm}
  $ \fra{\delta^2 \p_s\,\gk}{\delta\tp(p,\nu) \delta \psi(-p,-\nu)} = \fra{1}{2} \, \dtt$
\end{minipage}
\begin{minipage}[t]{5cm}
\includegraphics[width=50mm,height=15mm]{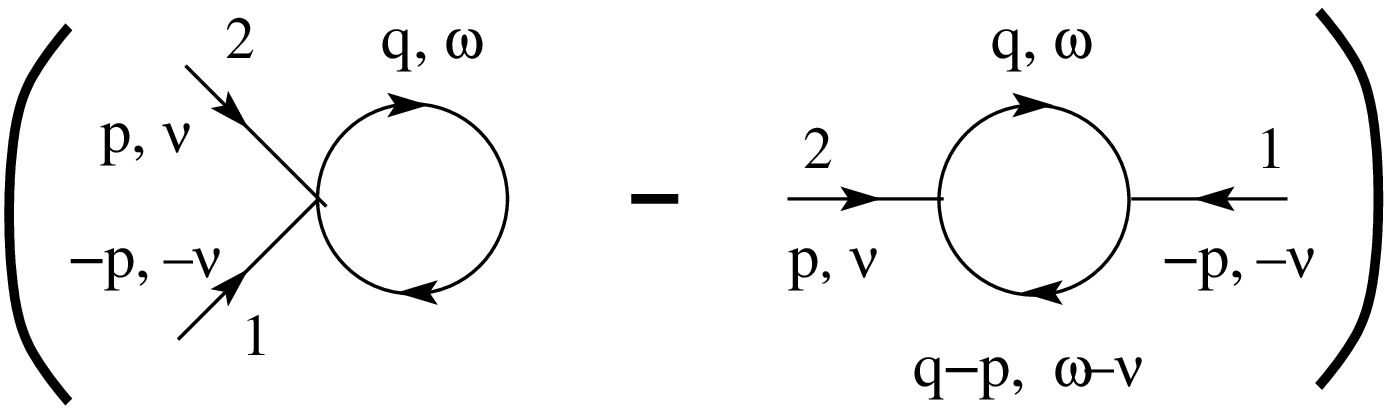}.
\end{minipage}
\end{center}
 In these graphs, the index 1 or 2 on external legs refers to
 the corresponding field ($\psi$ or $\tp$ respectively),
  the $n$-point vertices correspond to $2\times2$ 
 matrices of   $\gk^{(n)}$ with  $n-2$ (external) fixed indices and 
  two summed over.  The 
  propagator lines stand for $[\hat{\Gamma}_k^{(2)} + \hat{R}_k]^{-1}$, 
 which can be easily computed from (\ref{anz}) and (\ref{cutoffmat}).   
 We don't detail the full computation of these graphs nor
  the subsequent explicit frequency integration which are lengthy but straigthforward.

Before giving  the resulting flow
 equations, let us put them in a suitable form for the search 
 of fixed point --
 since we are interested {\it in fine} in the scale invariant regime.
 First it is convenient to explicitely express the Ising symmetry 
 by  defining the functions $\tilde{U}_k(\rho) \equiv  U_k(\psi)$,
 $\tilde{Z}_k(\rho)\equiv Z_k(\psi)$ where $\rho=\psi^2/2$ is the $Z_2$ invariant.
  The derivatives of the functions of $\tilde{U}_k(\rho)$ and $U_k(\psi)$
 are simply related: $\p_\psi U_k = \psi \p_\rho U_k$, $\p_\psi^2 U_k=  \p_\rho U_k + 2\rho  \p_\rho^2 U_k$ 
 $\dots$ and similarly for $Z_k$ and $\tilde{Z}_k$. As for the flow equation,  $\p_s (\p_\rho \tilde{U}_k) = 1/\psi\p_s (\p_\psi U_k)$
 for nonzero $\psi$ and $\p_s \tilde{Z}_k = \p_s Z_k$.
 Then to absorb any explicit dependence on the running
 scale $k$, we  introduce the dimensionless
 quantities   (according to (\ref{anz})) 
\begin{equation}
\left\{
\begin{array}{r l l}
\bp  & = &k^{{2-d}} \,\dimz\, \rho \\
u(\bp) & = &k^{-d}  \,\tilde{U}_k(\rho)\\
z(\bp) & = & \dimz^{-1} \tilde{Z}_k(\rho)
\end{array}
\right. \label{defvaradim}
\end{equation}
where the subscript $k$ has been dropped on  dimensionless functions.   
We further introduce a dimensionless cutoff function  
 $r(y) = Q_k(q^2)/(\dimz q^2)$ where $y=q^2/k^2$ and  hence 
 $\p_s Q_k = \dimz\, k^2\, s(y)$ with
 $s(y)=-\eta_Z\,y\,r(y)-2\,y^2\,\p_y r(y)$.
Finally, the flow equations for the dimensionless 
 functions $u'$ and $z$ are given by:
\begin{equation}
 \begin{array}{r c l}
\p_s u'&=&u'\,\left( -2 + \eta_Z \right)
      +\left( -2 + d + \eta_Z \right) \bp \,u'' + \displaystyle\frac{1}{2}\, \left(3\,u'' + 2\,\bp\,u''' \right)\, L_1^d + \displaystyle\frac{z'}{2}\,L_1^{2 + d} \\
\p_s z&=&z\,\eta_Z  +\left( -2 + d + \eta_Z  \right) \bp \,z' + 
\displaystyle\frac{1}{2}\, \left(z' + 2\,\bp\,z'' \right)\, L_1^{d} - 2\,\bp\,z'\,
   \left( 3\,u'' + 2\,\bp\,u''' \right)\,L_2^{d}\\
 &+& \displaystyle\frac{1}{d}\left\{-\left( 1 + 2\,d \right) \,\bp\,
     {z'}^2 \,L_2^{2 + d}
   +2\,\bp\, {\left( 3\,u'' + 2\,\bp\,u''' \right) }^2 \,M_4^{d}\right.\\
&+& \left. 4\,\bp\,z'\,
     \left( 3\,u'' + 2\,\bp\,u''' \right)\,M_4^{2 + d} +2\,\bp\,{z'}^2\, M_4^{4 + d}
\right\}\\
 \p_s \ln X_k &=&  \displaystyle\frac{1}{2}\bp\,
     {\left( 3\,u'' + 2\,\bp\,u''' \right) }^2\,L_3^{d} + 
  \bp\,z'\,
   \left( 3\,u'' + 2\,\bp\,u''' \right)\,L_3^{2 + d} + \displaystyle\frac{1}{2} \bp\,{z'}^2\,L_3^{4 + d}
\label{dsx}
\end{array}
\end{equation}
where primes denote derivatives wrt $\bp$ and 
the so-called thresholds funtions $L$ and $M$ are defined by:
\begin{equation}
\begin{array}{l c l}
L_n^d &=& -(n+\delta_{n0})\,v_d\, \displaystyle\int dy\,y^{d/2-1}\,
 \displaystyle\frac{s(y)}{h(y)^{n+1}}\\
M_n^d &=& v_d\, \displaystyle\int dy\,y^{d/2}\,
 \left(\displaystyle\frac{-(n+\delta_{n0})\,s(y)\,\left(\p_y h(y)\right)^2}{h(y)^{n+1}}+
 \displaystyle\frac{2\,\p_y s(y)\,\p_y h(y)}{h(y)^{n}}\right)
\end{array}
\end{equation}
with $v_d^{-1}= 2^{d}\pi^{d/2}\Gamma(d/2)$ and
$h(y)=y(z+r(y))+ u'+2\,\bp\,u''$.
By definition (see Section \ref{secAnz}), the anomalous dimension  $\eta_X\equiv -\p_s \ln \dimx$ 
  is obtained  by evaluating Eq. (\ref{dsx}) at the running minimum $\psi_0$
 of the  potential $U_k$,  or equivalently at the minimum $\bp_0$ 
 of $u$.
 Similarly,  $\eta_Z$ is obtained by solving at the minimum
  the equation $\p_s z |_{\bp_0} + z' \,\p_s \bp_0 =0$ where $\p_s z$ is given by (\ref{dsx}).
  The additional contribution $z' \,\p_s \bp_0$ is generated by the running of 
 the minimum implicitely defined as $u'(\bp_0)=0$.
 Indeed, the running of  $u$ implies that its minimum flows
   according to  $\p_s u'(\bp_0) = 0 = \p_s u'|_{\bp_0} + \p_s \bp_0 \,u''|_{\bp_0}$
 which, using (\ref{dsx}) evaluated at $\bp_0$  yields the expression for $\p_s \bp_0$.

 We emphasize that, as is to be expected from FDT, the dynamics decouples 
 from the statics, {\it i.e.} the  \npt flow equations (\ref{dsx}) 
  for $u'$ and $z$ do not depend on $X_k$ (and they identify with  those derived in equilibrium 
 with the \anz (\ref{ising})  \cite{berges02}).
Note that  the threshold functions $L$ and $M$ intervening 
 in the equation for the anomalous dimensions
 can be computed analytically upon the choice of the  cutoff
 $r(y)=(1/y-1)\,\theta(1-y)$  (corresponding to  
 (\ref{litim})), which greatly simplifies the numerical resolution.
 The numerical procedure to integrate the flow equations (\ref{dsx})
  is detailed in the next section.

\section{Numerical integration of the flow equations}
\label{secNum}

 We will consider different levels of approximation.
 In a first step  --- which will be referred to as local potential
  approximation (LPA) --- 
 the field dependence of the kinetic renormalization function $z$
 can be neglected,
 {\it i.e.}  only  a running coefficient $\dimz$ 
 is considered.
 When restoring the $\bp$-dependence of $z$,  
  the corresponding approximation will be denoted UZA.

 For both approximations, the numerical procedure to determine
  the fixed point solution of Eq. (\ref{dsx})
 and to compute the critical exponents is  the following.
  We sample the field $\bp$ on a mesh of spacing $\Delta \bp$ and  
 discretize the flow equations (\ref{dsx}) using finite differences 
  at order $\Delta \bp^4$ to calculate the $\bp$-derivatives of $u$ (and $z$).
 For the integrals, we either use their analytical expression
(whenever available) or calculate them numerically using Simpson's rule.
  We implement an explicit forward integration scheme to propagate the solution
 between scale $s$ and $s+ \Delta s$,
   which turns out to be stable for sufficiently small $\Delta \bp$ and $\Delta s$.
 The convergence of the numerical procedure when varying $\Delta \bp$ and $\Delta s$
   has been carefully checked.
  We start out at the microscopic  scale $s=0$ ($k=\Lambda$) from a quartic bare potential
  $u(\bp) = \lambda/2(\bp-\bp_\Lambda)^2$ where $\bp_\Lambda$ represents  the temperature. 
  We carry through the numerical integration by lowering $s$ towards 
 $s\to-\infty$. For large bare $\bp_\Lambda$, the system
  flows to the symmetric (high temperature) phase
 where the (dimensionfull renormalized) minimum of the potential  $\rho_0=k^{d-2}\dimz^{-1} \bp_0$ 
 vanishes,  whereas for small $\bp_\Lambda$, it flows to the broken
 (low temperature) phase where $\rho_0$ acquires a finite value as $s\to -\infty$.
  For  a fine-tuned 
  initial  $\bp_\Lambda^c$, the system is in the critical regime, which corresponds 
 to the effective potential $u$ (and $z$) flowing to a fixed point (scale invariant) form $u^*$ (and $z^*$). 
 The critical exponents $\eta$ and $z$ can then be computed from
  the fixed point values
 of $\eta_Z$ and $\eta_X$. The critical index $\nu$ is obtained by linearizing the flow
  in the vicinity of $(u^*,z^*)$ and determining the  (negative) eigenvalue 
 characterizing  the unstable (relevant)  direction.
 This procedure is carried out within the LPA and UZA approximations.
  The results are gathered in tables \ref{tab1} and \ref{tab2}  and are commented  in the next section.

\section{Results}
\label{secRes}

The critical exponents for Model A  obtained in this work from the NPRG equations (\ref{dsx})
 are summarized in Tables \ref{tab1} and  \ref{tab2} for  dimensions $d=3$ and $d=2$
 respectively, and compared with results ensuing from other field theoretical methods and Monte Carlo
 simulations. 

Let us first comment on the equilibrium exponents $\eta$ and $\nu$.
 As emphasized in Section \ref{secFlow}, the dynamics  decouples from the statics 
  in Eqs. (\ref{dsx}) --- as expected from TRS in the stationary regime.
  As a consequence, the equilibrium exponents computed here
  should match (up to numerical accuracy)  those obtained in earlier NPRG works
  on the (equilibrium) Ising model. The three-dimensional Ising model 
 has been thoroughly investigated within the NPRG framework as a testing ground of the method \cite{bagnuls01}. 
 In particular, critical exponents have been 
 computed  using the  cutoff function (\ref{litim}) in \cite{canet03b,ballhausen04} (though
 with different numerical procedures). The exponents $\nu$ and $\eta$ we obtain in $d=3$ 
 precisely  reproduce these values both at LPA and UZA.
 
 We know from these previous studies that the accuracy  can be improved
 by optimizing the  choice of the cutoff function.
 At order $\nabla^2$  in derivatives, optimized exponents  are 
 $\nu=0.6281$ and $\eta=0.0443$ \cite{canet03a}, which are already in close agreement with the 
 6-loop calculations \cite{guida98} or Monte Carlo simulations \cite{hasenbusch01}.
  However, since the determination of $\eta$ is related to the momemtum  structure
 of the two-point correlation function,  its accuracy is poorer than that of $\nu$.
 A better accuracy on $\eta$ requires to compute the next order $\nabla^4$
 in derivatives which yields $\eta=0.033$ \cite{canet03b}.
 In two dimensions, far fewer NPRG results are available. A  calculation
 with  cutoff function (\ref{litim}) has been achieved in \cite{ballhausen04} and both
 results are in close agreement.  As for in $d=3$, the determination
 of $\eta$ remains poorer than  that for $\nu$ at  order  $\nabla^2$ in derivatives.
  However, the two-dimensional Ising model has not been systematically investigated 
  and neither optimized nor order $\nabla^4$ exponents have  been determined in $d=2$. 

 We can now come to the new part of this work which concerns  the dynamics. 
 The situation for $z$ is very different from that of the equilibrium 
  critical exponents.  
 For the dynamics,  no high-loop expansions  or exact results 
 in $d=2$ are available. Furthermore, results from MC simulations appear to be 
 rather  scattered especially in $d=2$.
  The values reported in Tables \ref{tab1} and \ref{tab2}
 seem to be  accepted as  reference values \cite{calabrese05}.
  On the field-theoretical side, the determination of $z$ is very sensitive
 to the choice of the resummation scheme since only a few orders are known. 
  Various resummation schemes have been studied and  we quote here 
 the latest results obtained in \cite{prudnikov06}.

 Our results are in reasonable agreement with these various estimates.
 This is one of the key point of the NPRG approach and a central motivation for this work: 
 the leading order in derivatives appears to already provide a reliable determination 
 of physical quantities,  as outlined above. 
 Notice that the variation on $z$ between LPA and UZA is not meaningful
 and merely provides an estimate of the error. The reason is that
  going from LPA to UZA does not amount to enriching the \anz
 for the dynamical part since in both cases only a running coefficient $X_k$
 is allowed by TRS. Hence even at UZA, the accuracy on $z$
 remains poorer than that for the equilibrium exponents and the rapidity of convergence
 on $z$ can not be tested within these approximations. 
 One would need to implement  the next order in time 
 derivatives to improve the accuracy on $z$, which is rather costly.
  Alternatively,  one could modify the regularization scheme and 
 resort to  a frequency and noise coarse-graining as mentioned in Section \ref{secAnz},
 which is likely to yield better results, but is yet to be investigated.


\begin{table}
\begin{tabular}{|l|c|c|c|l|}
\hline
 ${d=3}$       &  $\nu$ & $\eta$ & $z$ & Ref.\\
\hline
 LPA     & 0.65  & 0.11 & 2.05 & [this work] \\
  UZA     &  0.63 &   0.05     & 2.14 &  [this work]\\
\hline
FT & 0.6304(13) & 0.0335(25)& &\cite{guida98}\\
MC     & 0.6297(5) & 0.0362(8)& &  \cite{hasenbusch01} \\
\hline
FT  & &     &  2.0237(55)&  \cite{prudnikov06}\\
MC      & & &  2.032(4)  2.055(10)& \cite{ito00,grassberger95} \\
\hline
\end{tabular}
\caption{Critical exponents of Model A in $d=3$ from the different 
 NPRG approximations (LPA and UZA) computed in this work, 
 compared with results from  field theoretical
 methods (FT) and Monte Carlo simulations (MC).}
\label{tab1}
\end{table}

\vspace{1cm}

\begin{table}
\begin{tabular}{|l|c|c|c|l|}
\hline
     ${d=2}$    &  $\nu$ & $\eta$ & $z$ &\\
\hline
 LPA     & 0.78 & 0.43 & 2.15 & [this work]\\
UZA       &  1.1     &   0.37     & 2.17 & [this work]\\
\hline
exact    & 1  & 1/4 & &\\
\hline
FT  & &     & 2.0842(39)&  \cite{prudnikov06} \\ 
MC      & & & 2.1667(5) & \cite{nightingale00}\\
\hline
\end{tabular}
\caption{Critical exponents of Model A in $d=2$ from the different 
 NPRG approximations (LPA and UZA) computed in this work, 
 compared with results from  field theoretical
 methods (FT) and Monte Carlo simulations (MC).}
\label{tab2}
\end{table}

\section{Summary}
\label{secConcl}

In this work, we have studied the critical dynamics of Model A
 within  the NPRG formalism.
   We have in particular detailed the 
 device of an appropriate approximation scheme preserving the symmetries,
  the derivation of the NPRG flow equations  and their resolution. 
  Using a very simple \anz, that is at the leading order in derivatives, 
  we have obtained a reliable estimate for the dynamical exponent $z$: 
 $z=2.09(4)$ in $d=3$ and $z=2.16(1)$ in $d=2$.
 The fact that the leading order already
 yields quantitative results  is a  generic feature of the NPRG approach \cite{berges02} which makes 
 it particularly powerful. 
 This feature is the central motivation
 for the emphasis put throughout this work on the methodological part:
   a leading order
 NPRG calculation  can already enable one to investigate nontrivial problems.  
 Restricting to \nequ critical phenomena, NPRG studies
  have indeed brought out new non-perturbative
  properties of reaction-diffusion processes   \cite{canet05} and 
 allowed to tackle interface growth problems \cite{canet06}.
 We hence believe
  the approach to be useful to investigate many other \nequ scaling phenomena. 
 Of course,   as the application of NPRG  techniques
  to \nequ statistical physics and dynamics is very recent, 
   a great deal of systematic studies remain to be done as well to test
   the efficiency of the different  \anz and regularisation schemes out-of-equilibrium, 
 which will be the goal of  future works.

The authors are indebted to B. Delamotte for fruitful exchanges 
 all through this work. The authors also wish to thank 
   G. Biroli, A. Lef\`evre  and N. Wschebor for enlightening discussions.

\section{Appendix: NPRG flow equation for $\gk$}
\label{app}

This appendix is dedicated to the derivation of the exact flow equation  (\ref{dkgam})
 for $\gk$.
 (A similar derivation  can be found for equilibrium systems in \cite{berges02}).
 In this appendix, we use the shorthand ${\xg}\equiv(x,t)$,
  vectors are denoted by capital letters ({\it eg.} $\Psi\equiv[\psi,\tp]$, ${\cal J}=[j, \tJ]$)
 and for functional derivatives we introduce the  notation
 $\dJ{\xg} \equiv [\delta/\delta j(\xg), \delta/\delta \tJ(\xg)]$
(and similarly for  $\dP{\xg}$). 
  Overhead  hat symbols are used for $(2\times2)$ matrices.

The variation of $\gk$ (at fixed $\Psi$)
  under an infinitesimal change of the running scale $k$ follows from Eq. (\ref{legendre}):
\begin{equation}
\pk \gk[\Psi_k]\big|_\Psi =- \pk \clw[{\cal J}]\big|_\Psi + \int_{\xg} \,\pk{\cal J}(\xg)\, \sca \,^t\Psi_k(\xg) - 
\demi \int_{\xg,\xg'} \Psi(\xg)\,\sca\,\pk \rk(\xg,\xg')\,\sca\,^t\Psi(\xg') \hbox{\hspace{0.5cm}} \label{exprclg}
\end{equation} 
 where $\clw\equiv \ln {\cal Z}_k$.
 The variation with $k$ of $\clw$ at fixed  
 $\Psi$ is related to that of $\clw$ at fixed ${\cal J}$ by: 
\begin{equation}
\pk \clw  \Big|_\Psi = \pk \clw  \Big|_{\cal J} + \int_{\yg} \pk {\cal J}(\yg) \sca \,^t\dJ{\yg}\clw. \label{exprclw1}
\end{equation} 
 The expression of  $\pk \clw  \big|_{\cal J}$ is obtained 
 by taking the derivative  of  (\ref{calz}) wrt $k$:   
\begin{equation}
\begin{array}{r c l}
\pk \ln {\cal Z}_k \Big|_{\cal J} &=& \displaystyle\int {\cal D}\Phi\, \Big\{-\fra{1}{2} \int_{\xg,\xg'} \Phi({\xg})\,\sca\, \pk \rk(\xg,\xg') \,\sca\, ^t\Phi({\xg'})\Big\}\, \expo{-{\cal S}_k - \Delta {\cal S}_k + ^t\Phi {\cal J}}\nonumber\\
 &=& \Big\{-\fra{1}{2} \int_{\xg,\xg'} \dJ{\xg}\,\sca\, \pk \rk(\xg,\xg') \,\sca\, ^t\dJ{\xg'} \Big\}\,\expo{\clw}\nonumber\\
 &=& \pk \clw \  \expo{\clw}.
\end{array}
\end{equation} 
After expressing the derivatives  $\dJ{}$ and $^t\dJ{}$ of  $\exp{(\clw)}$ and dividing out by $\exp{(\clw)}$ one obtains:
\begin{equation}  
\pk \clw \Big|_{\cal J} =   - \fra{1}{2} \int_{\xg,\xg'} \left\{ \dJ{\xg}\clw \,\sca \,\pk \rk(\xg,\xg')\,\sca \,^t\dJ{\xg'}\clw
  +  \hbox{Tr} \, \left[ \pk \rk(\xg,\xg')\,\sca\, \dJ{\xg} \big(\,^t\dJ{\xg'}\clw\big)\right] \right\}. \label{exprclw2}
\end{equation}
The last term in the right hand side   is the matrix of second (functional) derivatives of 
  $\clw$  which we denote $\derw(\xg,\xg')$.
 The flow equation of $\gk$ follows from inserting (\ref{exprclw1}) and (\ref{exprclw2})
  in~(\ref{exprclg}), which yields
\begin{equation}
\pk \gk[\Psi]\big|_\Psi = \demi \hbox{Tr}\,\int_{\xg,\xg'}\,  \pk\rk(\xg,\xg')\,\sca\,\derw(\xg,\xg').\label{exprclg2}
\end{equation} 

This equation can be conveniently expressed in a closed form upon  inverting $\derw$.
 The inverse of $\derw$ can be obtained by taking a functional derivative $^t\dP{\xg'}$ of
 the definition  $\Psi(\xg) = \dJ{\xg}\clw$: 
\begin{equation}
\widehat{1}\,\sca\,\delta^{(d+1)}(\xg,\xg') = \int_{\yg} \, ^t\dP{\xg'}{\cal J}(\yg)\,\sca\,\derw(\yg,\xg).
\end{equation} 
 The  matrix $^t\dP{\xg'}{\cal J}(\yg)$
 is simply given by  two successive derivatives  $\dP{\yg}$ and $^t\dP{\xg'}$
 of Eq~(\ref{legendre}), which yields~:  
\begin{equation}
^t\dP{\xg'} {\cal J}(\yg) = \derg(\xg',\yg) +\rk(\xg',\yg), \label{invgamespx}
\end{equation} 
 where  $\derg(\xg,\xg')$ denotes the matrix of second (functional) derivatives of  $\gk$. 
 Finally, inserting ~(\ref{invgamespx}) in the flow equation (\ref{exprclg2})
 yields the advocated equation: 
\begin{equation}
\pk \gk[\Psi_k]\big|_\Psi = \demi \hbox{Tr}\int_{\xg,\xg'} \,\pk \rk(\xg,\xg')\,
\sca\,\big[\derg+ \rk\big]^{-1}(\xg,\xg'),\label{dkgamdyn}
\end{equation} 
 which can be Fourier transformed and underlies the NPRG calculations of this work.


\end{document}